\documentstyle[11pt,newpasp,twoside,epsf]{article}

\begin{document}

\title{Tidally--induced angular momentum transport in disks}
\author{Caroline E.~J.~M.~L.~J. Terquem} 
\affil{Institut d'Astrophysique de Paris, 98bis Bd Arago,
75014 Paris, France}

\begin{abstract}
We discuss the transport of angular momentum induced by tidal effects
in a disk surrounding a star in a pre--main sequence binary system.
We consider the effect of both density and bending waves.  Although
tidal effects are important for truncating protostellar disks and for
determining their size, it is unlikely that tidally--induced angular
momentum transport plays a dominant role in the evolution of
protostellar disks.  Where the disk is magnetized, transport of
angular momentum is probably governed by MHD turbulence.  In a non
self--gravitating laminar disk, the amount of transport provided by
tidal waves is probably too small to account for the lifetime of
protostellar disks.  In addition, tidal effects tend to be localized
in the disk outer regions.


\end{abstract}

\section{Angular momentum exchange between the disk rotation 
and the orbital motion} 

In a binary system where at least one of the stars is surrounded by a
circumstellar disk, tidal waves excited by the companion propagate
into the disk.  If the disk and the orbital plane are coplanar, these
waves are called density waves.  In a noncoplanar system, both density
and bending waves are excited.  They are respectively of even and odd
symmetry with respect to reflection in the disk midplane.  The pattern
speed $\Omega_P$ with which the tidally excited pattern rotates is
$\omega$ and $2 \omega$ for density and bending waves, respectively,
where $\omega$ if the binary angular speed.  The disk is truncated by
tidal effects in such a way that its radius is not greater than about
one--third of the separation of the system (Papaloizou \& Pringle 1977,
Paczy\'nski 1977, Larwood et al. 1996 for the non coplanar case).
Therefore, $\Omega_P$ is smaller than the angular speed of the gas in
the disk, and the tidal pattern carries negative angular momentum (or,
in other words, the torque exerted on the disk by the companion is
negative).  Through dissipation of the waves, the disk then loses
angular momentum and disk material flows inwards, whereas the
companion star, which excites the waves, gains this angular momentum.

Since there is no corotation resonance in the disk, secular exchange
of angular momentum between the disk rotation and the orbital motion
occurs only if the waves dissipate, either in the bulk of the disk or
at its boundaries (Goldreich \& Nicholson 1989).  In a laminar disk,
dissipation of tidal waves may arise through shocks.  A shock front
forms when the group velocity of the wave relative to that of the
fluid (or, equivalently, the perturbed velocity) is supersonic.
Shocks are very dissipative so the wave amplitude cannot rise much
above the level where the front first forms.  In other words, the wave
is decelerated at the shock front in such a way as to restore
marginally sonic wave motion.  Conservation of wave action may tend to
cause the amplitude of the (shock--)wave to increase again as it
propagates further in, but this effect is balanced by shock
dissipation maintaining the amplitude at the marginal level.

Tidally--induced angular momentum transport has been considered as an
alternative to turbulent transport (Shu 1976, Sawada et al. 1986,
Spruit et al. 1987, Larson 1989, see also Larson in this volume).  It
is viewed as particularly attractive in disks where the ionization
level is too low for the magnetorotational instability to operate
(Balbus \& Hawley 1998).  However, while the presence of spiral waves
have been inferred in the accretion disk of the dwarf nova IP~Peg
(Steeghs et al. 1997), there are no indications that they are
associated with significant angular momentum transport.  Below we
consider successively the case of density and bending waves.

\section{Angular momentum transport by density waves}

There are two main obstacles to the propagation of density waves down
to small radii.  First, if the disk is vertically stratified, the wave
front tends to be tilted upwards so that the wave action migrates
towards the surface of the disk and into any atmosphere it possesses
where it can take on high amplitude and be dissipated (Lin et
al. 1990a, 1990b).  Only under the artificial assumptions of a strict
polytropic edge and no dissipation can it be channeled into and remain
in a very narrow surface waveguide (Ogilvie \& Lubow 1999).  Migration
of wave action towards the surface (or wave refraction) is more
effective for high azimuthal mode number $m$, but is still efficient
for the two--armed spirals which are predominantly excited in binaries.
This is because even though these waves have a larger wavelength in
the linear regime, if they become nonlinear their profile necessarily
distort and develop short wavelength components for which refraction
might be important.

The second obstacle to long range wave propagation is a low
temperature or, equivalently, a high Mach number (e.g., Spruit 1987,
Savonije et al. 1994, Godon et al. 1998, Armitage \& Murray 1998,
Blondin 2000).  This is because the characteristic wavelength of the
excited waves decreases with increasing Mach number, so that it
becomes very small compared with the scale associated with the forcing
potential (Lin \& Papaloizou 1993).  The torque, which is obtained by
integrating over the volume of the disk the perturbed mass density
times the tidal force, is then very small.

There have been a number of 3D numerical calculations of tidal shock
waves (see Yukawa et al. 1997 and references therein, Haraguchi et
al. 1999) but the loss of disk angular momentum has not been computed
in these calculations.  Spiral shocks were seen in some of these
simulations, but they were much less distinct than in 2D.  Even in 2D,
where refraction is absent and therefore tidal effects are
overestimated, the pattern observed in IP~Peg can be reproduced only
in the outer disk during outburst, when the enhanced viscosity pushes
the disk edge into a region of strong gravitational perturbations from
the secondary (Armitage \& Murray 1998), or for unrealistically hot
disks (Godon et al. 1998).  Note that observations themselves only
show spirality in the outer disk in IP~Peg.  In 2D, calculations by
Savonije et al. (1994) and Blondin (2000) suggest that wave--driven
accretion onto the central star occurs in disks where the Mach number
is smaller than about 10, whereas it is inefficient when the Mach
number is larger than about 20.  Since the Mach number in protostellar
disks is thought to be larger than 10 (the disk aspect ratio is around
0.05--0.1), tidally--induced transport is probably not significant in
the disk inner parts.

\section{Angular momentum transport by bending waves}

Bending waves are more efficient at transporting angular momentum in a
disk than density waves, because they have a longer wavelength
(Papaloizou \& Lin 1995).  For the same reason they can also propagate
down to smaller radii.

Papaloizou \& Terquem (1995) calculated the $m=1$ bending wave
response of an inviscid disk with the rotation axis misaligned with a
binary companion's orbital rotation axis.  They assumed that the waves
were dissipated by nonlinear interaction with the background flow
before reaching the disk inner edge, so that all their angular
momentum was deposited into the disk.  They found that $m=1$ bending
waves can lead to the accretion of the disk on a timescale in excess
of a few times $10^7$ years.

Terquem (1998) calculated the tidal torque in a viscous disk (where
the waves are viscously damped) and found that it can be comparable to
the torque communicated internally by horizontal viscous stress acting
on the background flow when the perturbed velocities in the disk are
on the order of the sound speed.  The tidal torque can exceed the
horizontal viscous torque only if the viscous stress tensor is
anisotropic with the parameter $\alpha$ which couples to the vertical
shear being larger than that coupled to the horizontal shear.  When
the perturbed velocities become supersonic, shocks reduce the
amplitude of the perturbation such that the disk moves back to a state
where these velocities are marginally sonic (Nelson \& Papaloizou
1998).  When shocks occur, the tidal torque exerted on the disk may
become larger than the horizontal viscous torque.  Terquem (1998) also
found that if the waves are reflected at the center, resonances occur
when the frequency of the tidal waves is equal to that of some free
normal global bending mode of the disk.  If such resonances exist,
tidal interactions may then be important even when the binary
separation is large.  However, it is unlikely that in a realistic
accretion disk waves can be reflected at the disk inner edge.
Therefore, in a viscous disk, it is unlikely that transport of angular
momentum is increased by more than a factor two or so by tidal
effects.  It was also found that if $\alpha$ is larger than about
$10^{-2}$, bending waves are damped before they can reach the disk
inner parts.

\section{Conclusion}

Tidal effects in pre--main sequence binary systems are important for
truncating protostellar disks and for determining their size.
However, once the disk is truncated, the calculations reviewed above
suggest that tidally--induced angular momentum transport does probably
not play a dominant role.  Where the disk is ionized enough so that
the magnetorotational instability can develop, transport of angular
momentum is most probably dominated by magnetic turbulence.  If the
disk is laminar, the amount of transport provided by tidal waves is
unlikely to be large enough to account for the dissipation of the disk
on a timescale on the order of a few million years.  In addition,
tidal effects tend to be exerted mainly at the disk outer edge, where
the perturbation is the strongest.  Strong tidal effects at the outer
edge of the disk would allow mass from the outer region to retract
inwards, subsequently weakening the tides at the edge.  Some pile up
of mass may result at smaller radii, but whether that would affect
significantly the whole disk would depend on the disk mass.

\begin{acknowledgements} 
  
I thank S. Balbus and J. Papaloizou for useful comments on an early
draft of this paper.

\end{acknowledgements}

\end{document}